\documentclass[10pt, conference]{IEEEtran}

\usepackage[colorlinks = true,
  linkcolor = blue,
  citecolor = blue,
  urlcolor = blue]{hyperref}

\usepackage{amsmath}
\usepackage{graphicx}
\usepackage{balance}
\usepackage{url}

\usepackage{fancyhdr} 
\fancypagestyle{firststyle}
{
\fancyhf{}
\fancyfoot[C]{\scriptsize{Proceedings of the IEEE 33rd International Requirements Engineering Conference Workshops (REW 2025), Valencia, IEEE, 2025, pp.~209--214. \\ This version is the authors' copy. The publisher's definite version is available online via \url{https://doi.org/10.1109/REW66121.2025.00033}.}}
}

\newcommand{\attck}{ATT\&CK\textsuperscript{\scriptsize\textregistered}}

\begin{document}

\title{An Alignment Between the CRA's Essential Requirements and the
  ATT\&CK\textsuperscript{\small\textregistered}'s Mitigations}

\author{
\IEEEauthorblockN{Jukka Ruohonen}
\IEEEauthorblockA{University of Southern Denmark \\
Email: juk@mmmi.sdu.dk}
\and
\IEEEauthorblockN{Eun-Young Kang}
\IEEEauthorblockA{University of Southern Denmark \\
Email: eyk@mmmi.sdu.dk}
\and
\IEEEauthorblockN{Qusai Ramadan}
\IEEEauthorblockA{University of Southern Denmark \\
Email: qura@mmmi.sdu.dk}
}

\maketitle

\begin{abstract}
The paper presents an alignment evaluation between the mitigations present in
the MITRE's \attck~framework and the essential cyber security requirements of
the recently introduced Cyber Resilience Act (CRA) in the European Union. In
overall, the two align well with each other. With respect to the CRA, there are
notable gaps only in terms of data minimization, data erasure, and vulnerability
coordination. In terms of the \attck~framework, gaps are present only in terms
of threat intelligence, training, out-of-band communication channels, and
residual risks. The evaluation presented contributes to narrowing of a common
disparity between law and technical frameworks.
\end{abstract}


\section{Introduction}

\thispagestyle{firststyle} 

The CRA~\cite{EU24a} was agreed upon in 2024. The enforcement will start in
2027. The CRA is a new product-specific cyber security law covering most
products with a network connection functionality; the notable exclusions are
medical devices, motor vehicles, ships and maritime equipment, and cloud
computing. Although stricter obligations are imposed upon products categorized
as important or critical, all products must comply with the CRA's essential
cyber security requirements.

To help in bridging a gap between legal and technical domains, the paper
investigates an alignment between the CRA's essential cyber security
requirements and the \attck~framework~\cite{MITRE25}. It is a comprehensive
empirical catalog of real-world adversarial tactics and techniques maintained by
the non-profit MITRE corporation. It was initiated in 2013. Although the
framework's focus is on advanced persistent threats, the mitigations offered in
the framework generalize to countering other threat actors as well. Perhaps
partially due to this generalizability, the \attck~framework has been
extensively used also in academic research~\text{\cite{AlSada24, JiangMeng25,
    Roy23}}. Among this research is also a previous work for using the framework
to elicit security requirements~\cite{Golushko20}. Another related work worth
explicitly mentioning is about the \attck's alignment with
standards~\cite{Kern24}. The paper aligns with and contributes to this line of
elicitation and evaluation research.

Recently, the alignment of the CRA's essential cyber security requirements has
been evaluated against the legal requirements imposed by the GDPR, that is, the
General Data Protection Regulation~\cite{Ruohonen25RE}. In addition, the CRA's
reporting obligations with respect to severe cyber security incidents have been
analyzed in conjunction with the EU's NIS2 directive for critical infrastructure
protection~\cite{Ruohonen25COSE}. There is also existing work on the CRA's other
reporting obligations with respect to vulnerability coordination and disclosure,
including the mandatory reporting of actively exploited
vulnerabilities~\cite{Ruohonen24IFIPSEC}. The present paper continues this
comparative evaluation work.

The alignment evaluation raises also the paper's practical relevance because the
regulation will be supported by standards. To this end, also European
standardization and other related organizations have already conducted
CRA-specific evaluation studies~\cite{ENISAJRC24, ETSI24}. Related work is being
done by open source software communities~\cite{ORCWG25} who too perceive
standards and standardized processes as important for reaching compliance with
the CRA~\cite{Lawson25}. While recognizing that the terms validation and
evaluation have specific meanings in requirements engineering~\cite{Wieringa06},
the evaluation term is used because the CRA's requirements and the \attck's
mitigations are not validated with an existing product or an implementation. It
should be also mentioned that the CRA's essential cyber security requirements
are legal requirements imposed by a law. Therefore, the paper can be framed also
toward existing requirements engineering research dealing with legal
requirements specifically~\cite{Ruohonen25RE, Hjerppe19RE, Netto19}.

A further important point is that the \attck's mitigations are categorized into
three groups: enterprise, mobile, and industrial control systems. Thus, not all
of the mitigations are strictly about products, which is what the CRA is mainly
about, but their generality still allows to evaluate the alignment. In fact,
many of the natural language descriptions for the mitigations resemble actual
requirements to some extent. They are also known as
countermeasures~\cite{Schumacher03} or security controls~\cite{Boyes24}. It
should be further emphasized that the CRA's essential cyber security
requirements are risk-based; they should be prioritized according to results
from risk analyses. Even though the \attck~framework could be used to help at a
risk analysis, as also demonstrated in the literature~\cite{Ahmed22}, the paper
only considers the alignment evaluation. Even under this restriction, the paper
helps at identifying relevant mitigations that can support compliance with the
CRA's essential requirements at a conceptual level, thus also bridging the gap
between legal and technical domains

The paper's remainder is structured into three sections. The opening
Section~\ref{sec: approach} elaborates the analytical evaluation methodology,
presenting also research questions to guide the evaluation. Then, the evaluation
results are presented in Section~\ref{sec: results}. The final Section~\ref{sec:
  conclusion} presents a concluding discussion.

\clearpage
\section{Approach}\label{sec: approach}

\begin{figure}[th!b]
\centering
\includegraphics[width=\linewidth, height=6.5cm]{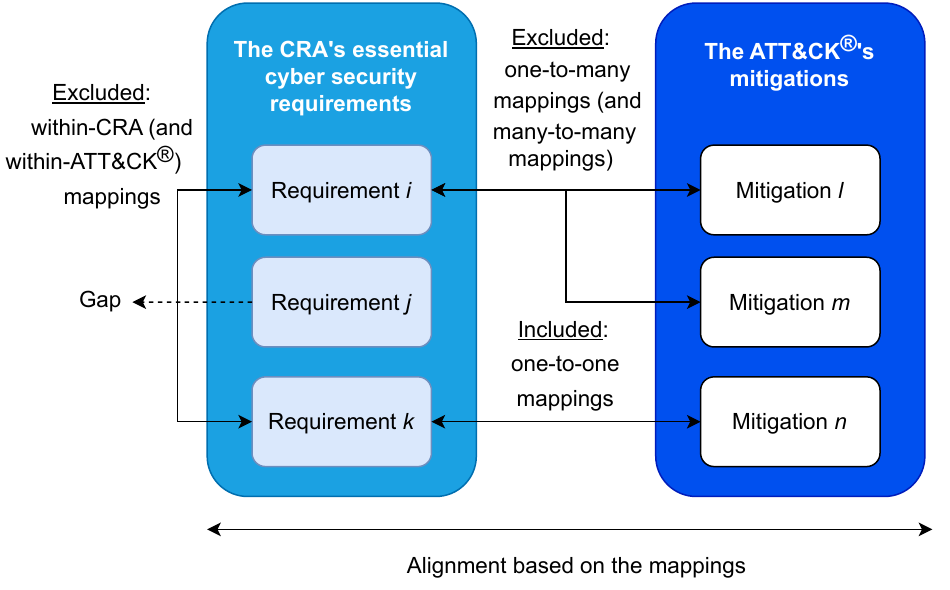}
\caption{Alignment, Gaps, and Mappings}
\label{fig: alignment}
\end{figure}

A terminological clarification is required to elaborate the analytical
evaluation methodology and the research questions~(RQs). The alignment is
evaluated by using a generic concept of a mapping. Different mappings have also
frequently been used to compare, evaluate, and develop standards and
frameworks~\cite{Roy23, Boyes24, Mussman20, Williams25}. Specifically, ``a
mapping is a bi-directional connection between information contained in two
artifacts'', whereas ``a link is a uni-directional connection from one artifact
to another'' \cite[p.~67]{Unterkalmsteiner15}. With this simple terminology,
which closely resembles definitions used in software traceability
research~\text{\cite{Jadoon19, Javed14}}, the artifacts examined refer to the
\attck's mitigations and the CRA's essential requirements categorized into
requirement groups in previous work~\cite{Ruohonen25RE}. A~bidirectional
connection between the two means that in case an essential cyber security
requirement is picked from the CRA, a corresponding mitigation is present in the
\attck~\text{framework---and} the other way around. To simplify the analysis, a
restriction is placed: each mitigation can only map to one essential cyber
security requirement. Thus, in terms of Fig.~\ref{fig: alignment}, only
one-to-one mappings are considered. In practice, a mapping of a mitigation was
done by considering the most representative and illuminating requirement.

Because the CRA is a law and the MITRE's \attck~is a framework, a prior
hypothesis is that at least some of the essential requirements map to multiple
mitigations because legal requirements cannot be specified in detailed,
technical terms. This point also correlates with research on the real or
perceived ambiguity of many legal requirements~\text{\cite{Netto19,
    Kempe24}}. Having said that, also the \attck~operates at a rather high
abstraction level, meaning that fine-grained technical details are absent from
the mitigations offered and suggested by the framework. In other words,
standards too are often ambiguous and incoherent. For instance, there is a lack
of consistency even regarding the definitions for fundamental concepts,
including for the confidentiality, integrity, and availability (CIA)
triad~\cite{Boyes24}. As the alignment evaluation is based on manually done
mappings, which have been the \textit{de~facto} approach in existing
work~\cite{Mussman20, Williams25}, the ambiguities and inconsistencies also
raise a validity threat, as is typical in qualitative research.

To address the threat, a collaboration process used in previous
work~\cite{Ruohonen25JSS} was adopted. The collaboration involved three steps:
(1) the first author made the initial mappings and identified the initial gaps;
(2) the other two authors reviewed these and raised arguments in case of
disagreements; (3) all three authors resolved the disagreements by short
negotiations. As usual, a perfect consensus holds for the final mappings and
gaps---after all, otherwise an author would not be an author.

Another point to draw from Fig.~\ref{fig: alignment} is that some of the
requirements may not map to any mitigations, or the other way around, which
would indicate a presence of a gap or several gaps. Explicit identification of
gaps in elicited requirements is generally important~\cite{Gaebert14}, including
with respect to changing requirements~\cite{Rolland04}, which can be seen to
frame also the CRA's essential cyber security
requirements~\cite{Ruohonen25RE}. With the previous points about legal
requirements in mind, a prior expectation again is that there are some gaps in
the CRA. On one hand: not everything can be covered in a law---and arguably not
everything should even be covered in a law. On the other hand: if there are
numerous gaps in the CRA \text{\textit{vis-\`a-vis}} the \attck, it could be
argued that the policy-makers overlooked some important aspects---a criticism
that has been expressed in relation to funding or other support for open source
software projects and communities~\cite{Ruohonen25JSS}. In any case, the overall
alignment can be summarized by counting the gaps, one-to-one mappings, and
one-to-many mappings.

With these elaborations, the following RQs are evaluated:
\begin{itemize}
\itemsep 3pt
\item{RQ.1: How well the CRA aligns with the \attck?}
\item{RQ.2: How many and what kinds of gaps there are?}
\end{itemize}

Before continuing to the results, a couple of additional remarks are warranted
about the manually but collaboratively done mappings. The first is that the
CRA's essential cyber security requirements were mapped to the mitigations by
using the twelve collated non-functional requirement groupings elicited and
conceptualized in previous work~\cite{Ruohonen25RE, Hjerppe19RE}. These simplify
the presentation because no explicit legal references are required. However, the
second point follows: the CRA's essential requirements contain also the CIA
triad, which can be argued to be related, either explicitly or implicitly, to
all other essential requirements. Initial disagreements were present
particularly with respect to the CIA triad's relation to the CRA's essential
cyber security requirement to review and apply exploitation mitigation
techniques. As said, these were resolved through negotiations between the
authors. A guiding principle for the negotiations was about the earlier remark
about mapping the mitigations to the ``most representative and illuminating
requirement''. To apply this guideline in practice, all authors were instructed
to first consider other requirements than the CIA triad, and then, if no
suitable candidate was present, to use the triad for a given \attck's
mitigation.

\begin{figure*}[p!]
\centering
\includegraphics[width=\linewidth, height=22cm]{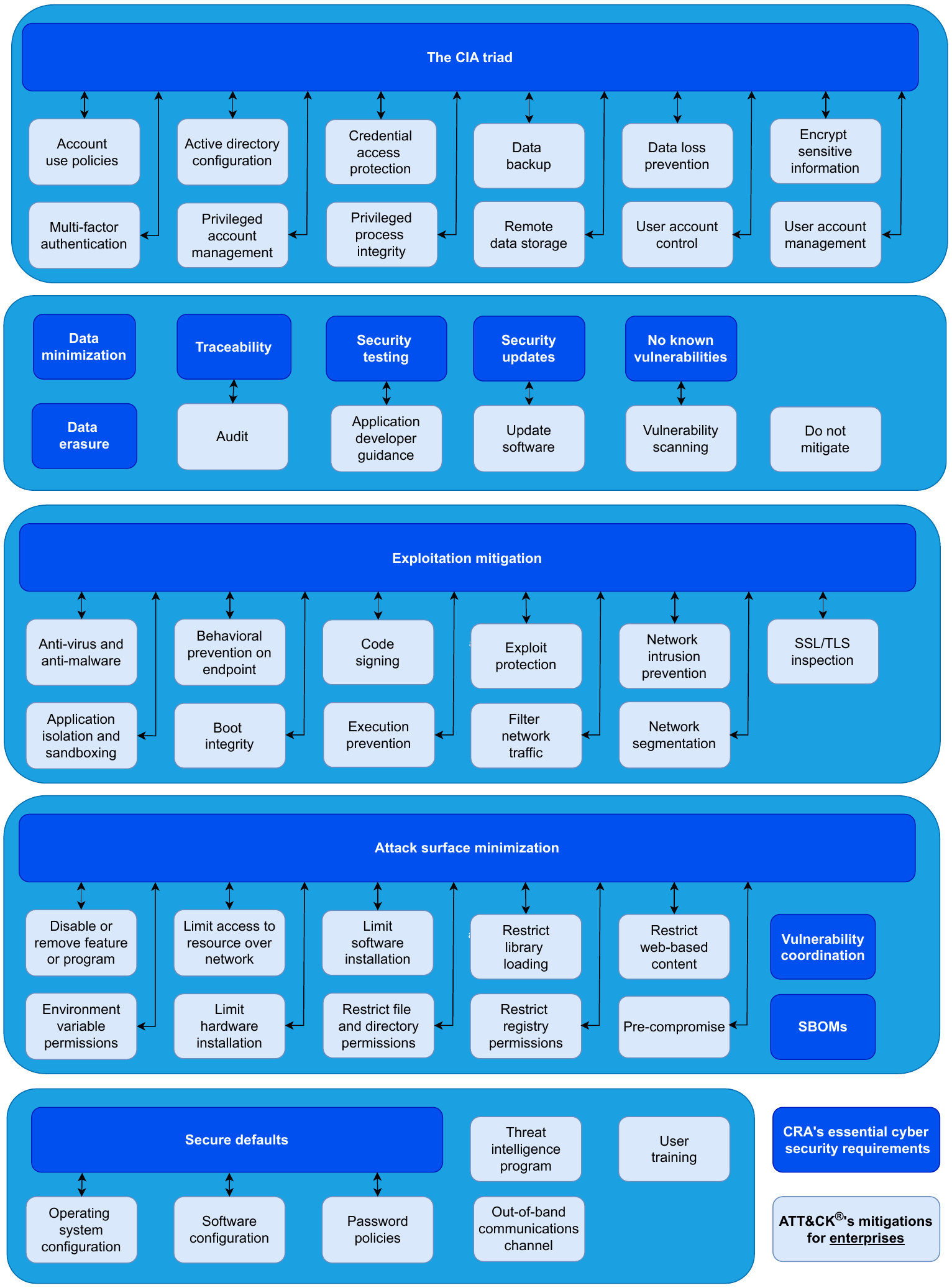}
\caption{The Mappings Between the CRA's Essential Cyber Security Requirements and the \attck's Mitigations for Enterprises}
\label{fig: mappings enterprises}
\end{figure*}

\begin{figure*}[p!]
\centering
\includegraphics[width=\linewidth, height=23cm]{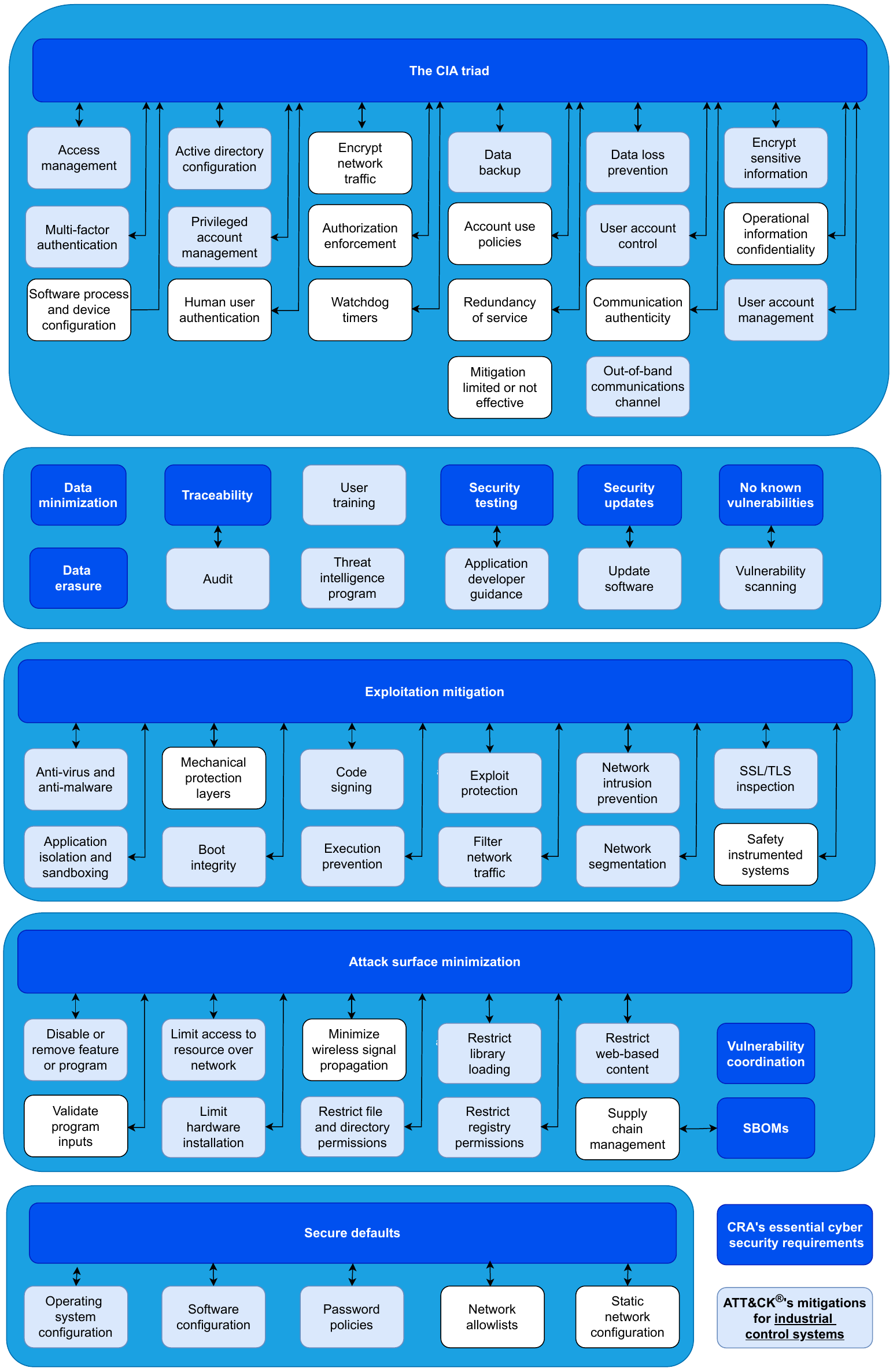}
\caption{The Mappings Between the CRA's Essential Cyber Security Requirements
  and the \attck's Mitigations for Industrial Control Systems
  (rectangles~colored in white denote those mitigations that are absent in
  Fig.~\ref{fig: mappings enterprises})}
\label{fig: mappings ics}
\end{figure*}

\section{Results}\label{sec: results}

The manually and collaboratively done mappings are shown in Figs.~\ref{fig:
  mappings enterprises} and \ref{fig: mappings ics} for the \attck's mitigations
for enterprises and industrial control systems, respectively. Before continuing
to unpack these mappings, a remark should be made: as the \attck~framework has
grown substantially throughout the years~\cite{AlSada24}, the mappings apply
only to the situation at the of writing, May 2025. With this point in mind, it
can be started by remarking that agreement between the three authors was very
good: in total, only four disagreements were recorded. In addition, five
disagreements were raised by the two reviewing authors about within-CRA and
within-\attck~mappings, which indicates that the exclusions in Fig.~\ref{fig:
  alignment} are too restrictive. As has already been pointed
out~\cite{Ruohonen25RE}, even the CRA's requirements are related to each other.

The main conclusion to draw from the two figures is that the alignment is
generally very good. Regarding the $12$ collated non-functional requirements of
the CRA and the $44$ mitigations for enterprises, only three mitigations could
not be mapped. These are the \attck's mitigations to have a threat intelligence
platform, to build an out-of-band communications channel for secure exchanges
during incident management, as also recommended by other
frameworks~\cite{NIST12}, and to provide training for users on matters such as
phishing and social engineering in general. All three gaps can be seen to be on
the organizational side. Also the CRA's essential requirements about security
testing and vulnerability coordination have been seen as being on an
organizational side instead of the technical
product-side~\cite{Ruohonen25RE}. Thus, in addition to the disagreements already
noted, some ambivalence is present regarding functional and non-functional but
technical requirements and organizational cyber security requirements.

Regarding the \attck's mitigations for industrial control systems, the gaps are
the same but with an addition of mitigations that cannot be reasonably
implemented. This ``mitigation limited or not effective'' category connotes with
a concept of residual risk in cyber security risk management research and
practice~\cite{NIST12, Khan22}. In other words, a \text{risk---as} a
probabilistic concept, still remains after a design and an implementation of
mitigations. Otherwise, the mitigations in Fig.~\ref{fig: mappings ics} are more
comprehensive than the mitigations for enterprises. The CIA triad stands out in
this regard. Unlike with the enterprise mitigations, also supply chain security
is accounted for. It maps to the CRA's obligation to establish a
machine-readable software bill of materials (SBOM). SBOMs are either required or
recommended also by other cyber security laws and frameworks~\cite{Williams25,
  Zahan23}. That said, the two mitigation categories are related also in a sense
that the CRA is envisioned to reduce supply chain management costs, risk
management costs, and incident management costs for users, including
enterprises, using the products covered by the law~\cite{ENISA24}. Also a few
other mitigations for industrial control systems can be explicitly
mentioned. Among these are recommendations to establish safety segmentation and
mechanical or physical protection layers, to minimize unnecessary wireless
signal propagation, to prefer predefined allowlists regarding network
destinations to which a device can connect, and to ensure availability by means
such as watchdog timers and redundancy solutions.

Finally, regarding the CRA's essential requirements, there are three visible
gaps: data minimization, data erasure, and vulnerability coordination. None of
these legal requirements can be seen to directly and explicitly map to any of
the mitigations. This observation supports an argument raised earlier in the
literature~\cite{Roy23} about a need to continuously update the
\attck~framework, including with respect to mitigations and
countermeasures. This updating point notwithstanding, as said, the framework
aligns well with the CRA in overall.

\section{Conclusion}\label{sec: conclusion}

The paper presented an alignment evaluation of the Cyber Resilience Act and the
\attck~framework's mitigations for enterprises and industrial control
systems. The conclusion is clear: the two align generally well with each other
(RQ.1). This conclusion can be seen to also raise the CRA's legitimacy in a
sense that its essential cyber security requirements are well-known means to
improve the cyber security of network-connected products. With respect to the
CRA, only data minimization, data erasure, and vulnerability coordination are
absent among the \attck's mitigations considered (RQ.2). With respect to the
mitigations, there are also some but not many gaps~(RQ.2). Among these gaps is a
recommendation related to residual risks. Since the CRA is a risk-based
regulation, meaning that also its essential cyber security requirements should
be prioritized according to risks identified, the residual risk concept is worth
explicitly singling out from the gaps.

Regarding future research, (a)~it seems sensible to further continue alignment
studies with alternative cyber security frameworks~\cite{NIST25} and
particularly standards. Such studies are important because following the
upcoming harmonized standards will in most cases provide a presumption of
conformity and compliance with the CRA, as is typical also in many other domains
in the European Union~\text{\cite{Hallensleben25, Ruohonen22ICLR}}. However,
alignment evaluations are not sufficient alone; (b)~once the standards have been
made, also they should be evaluated, including with respect to their practical
usefulness, preciseness and clarity~\cite{ETSI24}, rigor, scope, complexity, and
other related evaluation criteria for standards. In a similar vein, (c)~it seems
that empirical evaluations are mostly absent with respect to the \attck's
usefulness for practitioners. As was noted, the framework operates at a rather high abstraction level, which \text{may---or} may \text{not---decrease} its~practical~usefulness.

Regarding these three paths for future research, the concepts of layering and
layers could be used to move beyond mappings and alignments---a high-level
concept, such as a legal cyber security requirement, should ideally descend into
lower level layers, including concrete, technical
layers~\cite{Schumacher03}. This point also reiterates the importance of
standards, which too should arguably be specific enough but still applicable to
a wide range of products. Reaching a good balance in this regard may even be
crucial for a success of the CRA in improving the cyber security of
network-connected products in the future.

\balance
\bibliographystyle{ieeetr}

\end{document}